\documentclass[twocolumn,preprintnumbers,amsmath,amssymb]{revtex4}
\usepackage{epsfig}
\begin{document}

\title{Spin-transfer-induced excitations in bilayer magnetic nanopillars at high fields: \\ the effects of contact layers}

\author{Wenyu Chen$^1$, Andrew D. Kent$^1$, M. J. Rooks$^2$, N. Ruiz$^2$, Jonathan Z. Sun$^2$}
\affiliation{$^1$Department of Physics, New York University, New
York, NY 10003, USA} \affiliation{$^2$IBM T. J. Watson Research
Center, P.O. Box 218, Yorktown Heights, NY 10598, USA}
\date{October 6, 2005}

\begin{abstract}
Current-induced excitations in bilayer magnetic nanopillars have
been studied with large magnetic fields applied perpendicular to
the layers at low temperatures. Junctions investigated all have
Cu/Co/Cu/Co/Cu as core layer stacks. Two types of such junctions
are compared, one with the core stack sandwiched between Pt layers
(\emph{Type A}), the other with Pt only on one side of the stack
(\emph{Type B}). Transport measurements show these two types of
junctions have similar magnetoresistances and slope of critical
currents with respect to field, while  \emph{A} samples have
higher resistance. The high field bipolar excitation, as was
previously reported [\"{O}zyilmaz \emph{et al.} Phys. Rev. B,
\textbf{71}, 140403(R)] , is present in \emph{B} samples only.
This illustrates the importance of contact layers to
spin-current-induced phenomena. This also confirms a recent
prediction on such spin-wave excitations in bilayers.

\end{abstract}
\maketitle

One geometry to study spin-transfer-induced magnetization
excitation in magnetic nanopillars is to have a magnetic field
larger than demagnetization field $4\pi M_\text {s}$ applied
perpendicular to the current-perpendicular (CPP) spin valve layers
\cite{Ozyilmaz2003}. In this geometry, the in-plane shape
anisotropy of the nanopillar has only a small effect on the
magnetization dynamics and resulting phase diagram for
current-induced magnetic excitations. Recently, Polianski and
Brouwer \cite{Polianski2004} and Stiles \emph{et al.}
\cite{Stiles2004} predicted current-induced non-uniform spin wave
excitations even in \emph{single} thin ferromagnetic layer
nanopillars, provided that the pillar is asymmetric in the current
direction. \"{O}zyilmaz \emph{et al.} observed such excitations in
experiments on asymmetric single ferromagnetic layer pillar
junctions \cite{Ozyilmaz2004}. As predicted, excitations were
absent in more symmetric nanopillars \cite{Ozyilmaz2004}. In later
work, high field bipolar excitations were observed in bilayer
nanopillars and associated with non-uniform spin-wave excitations,
similar to those reported in single layers \cite{Ozyilmaz2005}. It
was recently shown theoretically that in bilayers, macrospin and
non-uniform excitations can compete, with the favored mode
depending on the device structure \cite{Brataas2005}. Here we
compare two types of magnetic nanopillars: one with two Pt layers
on both sides of an asymmetric Cu/Co/Cu/Co/Cu bilayer structure
(\emph{Type A}), and one with a Pt layer only on one side of this
structure (\emph{Type B}). Results are compared to theoretical
predictions \cite{Brataas2005}.

Using a nanostencil process \cite{Sun2002, Sun2003}, hundreds of
pillar junctions with submicron lateral dimension were fabricated
on a $1 $~cm $\times ~1 $~cm Si substrate. Stencil holes with
different but accurate lateral dimensions were opened up at the
depth of 75 nm, and pillar junctions were deposited through metal
evaporation. \emph{Type A} structures have Pt layers on both sides
of the core structure, \emph{i.e.} $\parallel 100 ~nm ~Cu\mid 15
~nm ~Pt_{70}Rh_{30}\mid 3 ~nm Pt\mid 10 ~nm ~Cu\mid t ~Co\mid 10
~nm ~Cu\mid 12 ~nm ~Co\mid 10 ~nm ~Cu\mid 3 ~nm ~Pt\mid 200 ~nm
~Cu\parallel$. \emph{Type B} samples have a Pt layer only on the
bottom of the magnetic layer sequence: $\parallel 100 ~nm ~Cu\mid
15 ~nm ~Pt_{70}Rh_{30}\mid 10 ~nm ~Cu \mid t ~Co\mid 10 ~nm
~Cu\mid 12 ~nm ~Co\mid 200 ~nm ~Cu\parallel$. We investigated
approximately 30 junctions of each type. On samples of each type,
junctions with $t \simeq 1.9$, $3.3$, and $4.3 ~nm$ and lateral
dimensions $50 ~nm \times 50 ~nm$ and $50 ~nm \times 100 ~nm$ were
studied in detail.

All measurements reported here were made at 4.2 K with a 4-point
geometry. Both DC resistance $V/I$ and differential resistance
$dV/dI$ for each junction were measured. A $0.2 ~mA$ AC current at
802 Hz was added to the DC current bias. Junction resistances were
found to scale inversely with their lateral areas. Positive
currents are defined such that electrons flow from thin Co layer
to thick Co layer.

Magnetoresistances (MR) were measured with magnetic field applied
in the film plane. Typical hysteresis loops are shown in the
insets of Fig. 1. These two junctions have identical $50 ~nm
\times 100 ~nm$ lateral areas and similar thin Co layer thickness
($t \simeq 3.3 ~nm$), with the loop of the \emph{Type A} in Fig
1(a) and that of \emph{Type B} in Fig 1(b). Due to the additional
Cu/Pt interface and Pt bulk scattering, junctions on sample of
\emph{Type A} are more resistive than those on sample of
\emph{Type B}. But their MR values are roughly the same, 2.2\% and
2.3\% for junction of \emph{Type A} and \emph{Type B}
respectively.

\begin{figure}[t]
\includegraphics[width=0.48\textwidth]{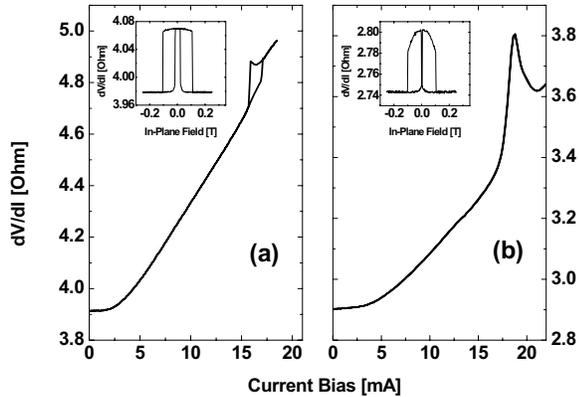}
\caption{Positive current sweep hysteresis loops of $50 ~nm \times
100 ~nm$ junctions with thin Co layer thickness $t \simeq 3.3 ~nm$
with a 7 T magnetic field applied perpendicular to the sample
surface at 4.2K. \emph{Type A} in (a) and \emph{Type B} in (b).
Insets: zero-current in-plane magnetoresistance hysteresis loops.}
\label{Singlesweep}
\end{figure}

$I(V)$ measurements were conducted with magnetic field applied
perpendicular to the sample surfaces. When the field is higher
than the demagnetization field $4\pi M_{\text {s}}$, the magnetic
layers are aligned along the field direction. When positive
current is applied to the junction, the spin transfer torque $a_J
\hat{m}\times (\hat{m}\times \hat{m_P})$ \cite{Slonczewski1996,
Berger1996} can drive the thin layer into instability. Here $a_J$
is the torque factor which is proportional to applied current
density $J$, and $\hat{m}$, $\hat{m_P}$ are the magnetic moment
unit vectors of the thin layer and the thick layer respectively.
When the torque is large enough, it may switch the thin layer into
the anti-parallel (AP) state. The differential resistance $dV/dI$
versus $I$ at 7 T is shown in Fig. 1, for the same two junctions
as in the insets. A resistance change close to in-plane MR was
observed from both curves. In Fig 1(a), the $I(V)$ curve of
\emph{Type A} junction shows a $\sim1.3 ~mA$ hysteresis and the
switching was accompanied by a step in $dV/dI$ in both directions
of current sweep. The majority of the junctions of \emph{Type A}
show hysteresis. However, most junctions of  \emph{Type B} (Fig
1(b)) do not have hysteresis, and their switching is accompanied
by a peak in $dV/dI$. Furthermore, DC resistance $V/I$ versus I
curves are smooth except that there is a sharp step at the
switching current. The step can be used to determine the critical
current.

In order to emphasize small features on top of the background
which is associated with Joule heating, the contour plots of
$d^2V/dI^2$ are shown in Fig. 2. Here the current is swept from
positive to negative with magnetic field fixed during each sweep.
Magnetic field decreases in steps of 0.05 T, from 7.5 T all the
way down to zero. (a) and (b) are the contour plots of $50
~nm\times 50 ~nm$ junctions with $t \simeq 1.9 ~nm$ for \emph{Type
A} and \emph{B} samples respectively. On both graphs, the
strongest feature corresponds to the switching between P and AP
state at positive currents. Above the demagnetization field
($\sim1.5 ~T$), the critical current increases with applied
magnetic field as shown in Fig. 2, which is consistent with
theoretical calculations of the threshold current for a macrospin
instability \cite{Slonczewski1996, Sun1999, Katine2000, Sun2000},
numerical modeling \cite{Ozyilmaz2003} and previous experimental
results \cite{Ozyilmaz2003}. The derivatives of critical current
density with respect to high field at 7 T $dJ_c/dH$ for these two
junctions are similar: $(3.1\pm0.5) \times 10^7 A/(cm^2 ~T)$ for
the one of \emph{type A}, and $(3.9\pm0.6) \times 10^7 A/(cm^2
~T)$ for that of \emph{type B}.

\begin{figure}[t]
\includegraphics[width=0.48\textwidth]{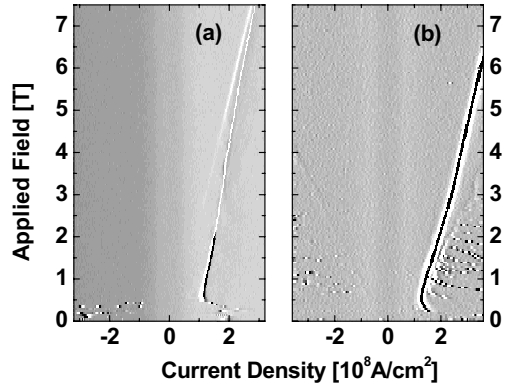}
\caption{$d^2V/dI^2$ current sweep contour plots of $50 ~nm \times
50 ~nm$ junction with $t \simeq 1.9 ~nm$ as the function of both
current density and magnetic field perpendicular to the sample
surface. (a): \emph{Type A} sample. (b): \emph{Type B} sample with
currents swept from positive to negative.} \label{Contourplot}
\end{figure}

In addition to the boundary of critical currents for switching,
there are more small features distributed in the contour plots
shown in Fig. 2. When positive current is applied to junctions of
\emph{Type A}, there is another boundary which is weaker than the
main switching peak and is located at slightly lower current bias.
Between these two boundaries, there are some even more complicated
structures. Those structures can even be seen on $dV/dI$ plots if
the part of the $dV/dI$ curve right before the switching is
magnified, like in Fig 3(a). Usually those excitations appear like
multiple peaks in $dV/dI$ for junctions of \emph{Type A}. However
the excitations in \emph{Type B} junctions at positive currents
behave differently. After switching, small features which look
like a ``comb" could be seen beyond a certain current boundary as
can be seen in Fig 2(b). Those excitations are multiple dips in
$dV/dI$ as shown in Fig 3(b). All junctions of  \emph{Type B}
studied exhibited this feature. The type of excitations in
junctions of \emph{Type B} were also studied in experiments before
\cite{Ozyilmaz2005}, while those in junctions of \emph{Type A}
were not.  All \emph{Type A} junctions studied do not show
additional excitations beyond the main peak in $dV/dI$ .

Furthermore, at negative currents, additional excitations are
found in Fig. 2(b). Such kind of excitations are absent below a
certain current boundary which is indicated as a weak line in the
contour plot of Fig. 2(b) and extends linearly to zero field zero
current. This was also revealed in the previous study
\cite{Ozyilmaz2005}. But in Fig. 2(a), no such excitations were
found at negative current polarity when $H>1.5 ~T$.

\begin{figure}[t]
\includegraphics[width=0.48\textwidth]{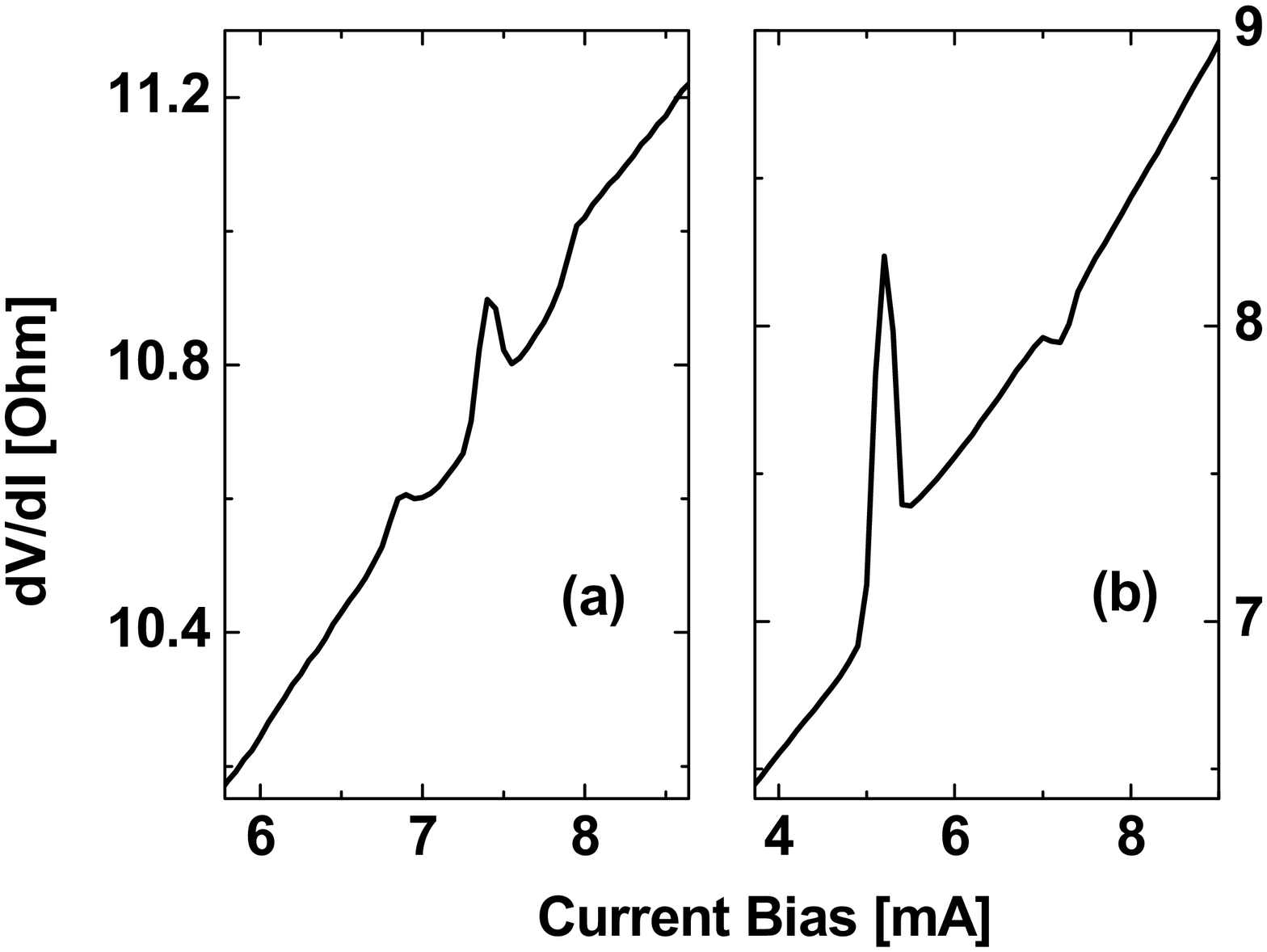}
\caption{$dV/dI$ current sweep curves of $50 ~nm \times 50 ~nm$
junctions with the regions of excitations at positive currents
magnified. (a): \emph{Type A} sample with $t \simeq 4.3 ~nm$, $H=
5 ~T$; (b): \emph{Type B} sample with $t \simeq 1.9 ~nm$, $H= 2.3
~T$} \label{Excitations}
\end{figure}

Theoretical studies of excitations in magnetic nanopillars go back
to Berger, who calculated an onset of spin wave excitations
\cite{Berger1996}, and Slonczewski, who considered a coherent
rotation of the whole magnet spin \cite{Slonczewski1996}.
Recently, Brataas \emph{et al.} \cite{Brataas2005} calculated the
onset of macrospin precession versus non-uniform spin-wave
excitations in bilayer magnetic nanopillars. Using the
two-spin-channel circuit theory, they deduced the spin torque on a
ferromagnet by considering its small transverse instabilities
$\delta\textbf{m}^{\bot}$ as the function of wavevector
$\textbf{q}$. Both the uniform macrospin ($\textbf{q} \rightarrow
0$) excitations and non-uniform spin-wave ($\textbf{q} \rightarrow
\infty$) excitations have been discussed. As an example, they
computed the phase diagram for bilayer junctions without Pt on
top. At negative currents, they predicted a spin-wave instability
for the \emph{thick} ferromagnet. At positive currents starting
from P state, there is a macrospin instability of the thin layer,
and after switching of the thin layer a further increase in
current leads to a spin-wave instability of the  \emph{thick}
layer. These predictions are consistent with our results in
junctions of \emph{Type B} and previous experimental work
\cite{Ozyilmaz2005}.

One of the results of \cite{Brataas2005} is that the layer
contacts are important in determining the resulting instabilities,
\emph{i.e.} whether it is a uniform or non-uniform mode and the
critical current. For simplicity, if the resistances of the two Co
layers are mainly due to the Cu/Co interfaces, then their
effective resistances in the circuit are close to each other. In
this case, the spin torque on the thin/thick magnetic layer in the
short wavelength limit can be written as:
\begin{eqnarray}
  \tau_{thin/thick}&=&\mp (P/2) \delta\textbf{m}_{thin/thick}^{\bot}  j^{(c)} \nonumber  \\
  &\times & (R_{m}-2R_{b/t})/R_{thin/thick}^{\uparrow \downarrow} \; \label{eq1}
\end{eqnarray}
where $P$ is the total polarization of the current, $R_{m},
~R_{b/t}$ are the resistances of normal metal contact layers
located between Co layers and outside Co layers close to
bottom/top respectively, $R_{thin/thick}^{\uparrow \downarrow}$ is
the mixing resistance of the thin/thick Co layer
\cite{Brataas2000} and $j^{(c)}$ is the charge current density in
the junction. This shows that the thickness of the top Cu contact
layer adjacent to thick Co affects the magnitude of the short
wavelength spin torque on the thick Co layer: increasing the
resistance asymmetry $|R_{m}-2R_{t}|$, increases the torque.

We applied this result to our  junctions. Parameters given by MSU
group \cite{Bass1999} were used to calculate Co resistances. For
Co layer with $t=3.3 ~nm$, the bulk contribution is 24\% of
interface counterpart. In this case, the thin layer resistance is
33\% lower than that of thick layer, and Eq \ref{eq1} is a
reasonable approximation. In \emph{Type A} samples, the Pt layers
on both sides of the pillar create a good zero spin accumulation
boundary so that the effective resistance of $R_{t}$ does not
extend over the Pt layers. In our \emph{Type A} junctions,
$|R_{m}-2R_{t}|\simeq R_m$. Whereas in \emph{Type B} samples the
top Cu lead is considerably longer (35 to 50 nm) and
$|R_{m}-2R_{t}| \gtrsim 6R_m$. In this case the short wavelength
spin torque is considerably larger and excitations should occur
for lower current densities. This is consistent with our
experimental results that show excitations at positive currents
beyond the main peak and negative currents only in \emph{Type B}
samples. Similar excitations in \emph{Type A} samples are
predicted to occur only at much larger current densities.

In summary, we compared spin transfer effects, especially the
current induced excitations, in bilayer magnetic nanopillars with
two types of layer stacks, one with Pt on both sides (\emph{Type
A}) and one with Pt only on one side (\emph{Type B}). In contrast
to the junctions of \emph{Type B}, no excitations were observed on
the junctions of \emph{Type A} both at positive currents higher
than switching threshold and negative currents, which confirms a
prediction of Ref \cite{Brataas2005}. In this sense, the contact
layers in bilayer magnetic nanopillars are of great importance in
spin-transfer devices.

This research is supported by NSF-DMR-0405620 and by ONR
N0014-02-1-0995.

\end{document}